\documentclass[aps,prl,twocolumn,superscriptaddress,amsmath,floatfix,11pt,reprint]{revtex4-2} 
\usepackage[utf8]{inputenc}
\usepackage[T1]{fontenc}
\usepackage{graphicx}  
\usepackage{dcolumn}   
\usepackage{bm}        
\usepackage{amssymb}   
\usepackage{framed}
\usepackage{amsmath}
\usepackage{multirow}
\usepackage{hhline}
\usepackage{color}
\usepackage{xcolor}
\definecolor{bluegreen}{rgb}{0,0.2,0.8}
\usepackage{subfigure,amsmath,verbatim,moreverb}
\usepackage{tabularx}
\usepackage{adjustbox}
\usepackage{lipsum}
\usepackage{longtable}
\usepackage{booktabs}
\usepackage{adjustbox}
\usepackage{hyperref}
\UseRawInputEncoding
\usepackage{graphicx}
\usepackage{booktabs}
\usepackage{siunitx}
\definecolor{lightcyan}{rgb}{0.88,1,1}

\usepackage{etoolbox}
\AtBeginEnvironment{align}{\setcounter{subeqn}{0}}
\newcounter{subeqn} %

\newcommand{\R}{\mathbf{r}}

\newcommand{\Eq}[1]{Eq.~(\ref{#1})}

\begin{document}

\title{
Simplified, Physically Motivated, and {\color{black} Broadly} Applicable Range-Separation Tuning
}
\author{Aditi Singh}
\email{aditisingh4812@doktorant.umk.pl}
\affiliation{Institute of Physics, Faculty of Physics, Astronomy and Informatics, Nicolaus Copernicus University in Toru\'n,
ul. Grudzi\k{a}dzka 5, 87-100 Toru\'n, Poland}
\author{Subrata Jana}
\email{subrata.niser@gmail.com, subrata.jana@umk.pl}
 \affiliation{Institute of Physics, Faculty of Physics, Astronomy and Informatics, Nicolaus Copernicus University in Toru\'n,
ul. Grudzi\k{a}dzka 5, 87-100 Toru\'n, Poland}
\author{Lucian A. Constantin}
\affiliation{Institute for Microelectronics and Microsystems (CNR-IMM), 73100 Lecce, Italy}
\author{Fabio Della Sala}
\affiliation{Institute for Microelectronics and Microsystems (CNR-IMM), 73100 Lecce, Italy}
\affiliation{Center for Biomolecular Nanotechnologies, Istituto Italiano di Tecnologia, 73010 Arnesano, LE, Italy}
\author{Prasanjit Samal}
\affiliation{School of Physical Sciences, National Institute of Science Education and Research, An OCC of Homi Bhabha National Institute, Bhubaneswar 752050, India}
\author{Szymon \'Smiga}
\email{szsmiga@fizyka.umk.pl}
\affiliation{Institute of Physics, Faculty of Physics, Astronomy and Informatics, Nicolaus Copernicus University in Toru\'n,
ul. Grudzi\k{a}dzka 5, 87-100 Toru\'n, Poland}

\date{\today}

\begin{abstract}
Range-separated hybrid functionals (RSH) with ``ionization energy'' and/or ``optimal tuning'' of the screening parameter have proven to be among the most practical and accurate approaches for describing excited-state properties across a wide range of systems, including condensed matter. However, this method typically requires multiple self-consistent calculations and can become computationally expensive and unstable, particularly for extended systems. In this work, we propose a very simple and efficient alternative approach to determine the screening parameter for RSH based solely on the total electron density of the system and the compressibility sum rule of density functional theory (DFT). This effective screening parameter achieves remarkable accuracy, particularly for charge-transfer excitations, surpassing the performance of previously suggested alternatives. Because it relies only on the electron density, the proposed approach is physically transparent and highly practical to automate DFT calculations in large and complex systems, including bulk solids, where ``tuning'' is not possible.
\end{abstract}

\maketitle

{\it{Introduction:}}~Since its advent, Density Functional Theory (DFT)~\cite{burke2012perspective} has become an indispensable formalism in interdisciplinary research, with significant applications in materials science and quantum chemistry~\cite{Jones2015,burke2012perspective,perdew2005prescription,Zhao2008Density}. Ground-state properties can often be predicted with reasonable accuracy using cost-efficient semilocal approximations such as the Local Density Approximation (LDA)~\cite{Perdew1992Accurate}, Generalized Gradient Approximations (GGAs)~\cite{perdewPRL96}, meta-GGAs~\cite{taoPRL03} or global hybrid\cite{becke1993density} level of approximations. In turn, the the excited-state properties within time-dependent DFT (TD-DFT), particularly Rydberg and charge transfer (CT) excitations, remain challenging~\cite{Casida-challenge,casida-challenge-2,casida-3,Zhao2006Density,RohrMartHerb2009,Hirata1999Time,Dreuw2003Long,Maitra2016Perspective,Knight2016Accurate,Richard2016Accurate,Ghosh2018Combining}. These limitations stem from the incorrect asymptotic decay of semilocal and global hybrid exchange-correlation potentials~\cite{Potential_Gilbert,Potential_issue,SmigaPot} (critical for Rydberg states) and the lack of long-range exchange (essential for CT excitations)~\cite{CTfails, CTfails2}, alongside the derivative discontinuity~\cite{kummel2008orbital,Godby1986} inherent in approximate density functionals. As an effective remedy, long-range corrected hybrid functionals with `ionization energy tuning' have been proposed by enforcing the exact, non-empirical Koopmans theorem (i.e., maintaining a constant for the long-range potential)~\cite{Baer2010Tuned,Stein2009Reliable,SteiEiseHele2010}. \textcolor{black}{Theoretically, the ionization potential-assisted tuning procedure optimizes the range-separation parameter $\omega$ to enforce the exact ionization energy (IE) condition. The resulting value,  $\omega_{IE}$, minimizes the expression}~\cite{kummel2018,Refaely2012Quasiparticle}
\begin{equation}
\omega_{IE} = \arg\min_{\omega} \left| IE(\omega) + \varepsilon_{\text{HOMO}}(\omega) \right|~.
\label{hy-tuned}
\end{equation}
Although this tuning procedure provides an enriching setting for small and medium-sized molecules,~\cite{Korzdorfer2012Assessment} repeated $\Delta$SCF calculations at the hybrid functional level become problematic. Consequently, it is very challenging to apply this scheme to periodic solids~\cite{kummel2018}, solvated or embedded systems~\cite{Srebro}, systems with strong non-covalent interactions~\cite{Refaely2012Quasiparticle}, large molecular chains, or nanostructured clusters~\cite{Karolewski2013Using}. As a potential substitute, schemes such as effective charge-transfer distance tuning~\cite{Yan2025Adaptable}, global density-dependent (GDD) tuning~\cite{Mandal2025Simplified}, and Electron Localization Function (ELF) tuning~\cite{Borpuzari2017new} have been proposed ({\textcolor{black}{for solids, we also recall Wannier-localization-based tuning}). These are one-shot (a black-box) strategies, 
that circumvent the need for a laborious scan over ionization energies (IEs), making them exceptionally beneficial for larger molecular systems. However, these are not universal, and their applications for periodic solids have never been explored. For solids and clusters, several procedures for determining range-seperated parameters have been developed that account for the distinct physical characteristics of extended systems~\cite{SkonGovoGall2014, Jana2023Simple, WeiGiaRigPas2018}. However, those approaches are generally not transferable to finite or molecular systems, particularly for accurately predicting ionization potentials or fundamental gaps~\cite{SkonGovoGall2014, Refaely2011Fundamental, Dahvyd2021Band}. Although all these methods offer valuable insights, there remains a strong need for simple and physically transparent procedures, especially for broader applicability for ``both-worlds'' molecules and solid-state physics. Simplified yet accurate tuning protocols are still lacking.

Thus, as a significant advancement, this Letter introduces a simple yet elegant alternative approach for determining the range-splitting parameter in screened hybrid functionals. The proposed formalism is conceptually straightforward and highly versatile, making it applicable to a broad range of systems in both quantum chemistry and solid-state physics. Its generality and ease of implementation offer a promising route for improving the accuracy and efficiency of electronic structure calculations across diverse fields. 


%
To establish the new formalism, we first recall the static density response function of the homogeneous electron gas (HEG), which can be conveniently represented as~\cite{giuliani2005quantum},
\begin{eqnarray}
 \chi(\mathbf{q}) = \frac{\chi_{KS}(\mathbf{q})}{1 - \left[v(\mathbf{q})
 +K_\textnormal{xc}(\mathbf{q})\right]\chi_{KS}(\mathbf{q})}~,
\label{eq:kernel}
\end{eqnarray}
where $\chi_{KS}(\mathbf{q})$ is the response in the KS framework, $v(\mathbf{q})=\frac{4\pi}{\mathbf{q}^2}$ is the conventional Coulomb potential, and $K_{xc}(\mathbf{q})$ is the static XC kernel (all representation is in reciprocal space with reciprocal space vector ${\bf{q}}={\bf{G}}-{\bf{G}}'$) given by~\cite{Constantin2016Simple,Constantin2007Simple}
\begin{equation}
 K_{xc}(\mathbf{q})=v(\mathbf{q})\left[\exp\left(-\frac{\mathbf{q}^2}{4\omega^2}\right)-1\right]~.
 \label{eq:kernel2}
\end{equation}
%

Here $\omega$ is the range separation (or screening) parameter that distinguishes between the short- and long-range components of the electron-electron interaction. In range-separated hybrid (RSH) functionals, this separation is often introduced via an Ewald-like decomposition, i.e.
\begin{eqnarray}
v(q) &=& \underbrace{v(q)\left[1 - \exp\left(-\frac{q^2}{4\omega^2}\right)\right]}_{\text{SR exchange}} + \underbrace{v(q)\exp\left(-\frac{q^2}{4\omega^2}\right)}_{\text{LR exchange}}~.
\label{eq1}
\end{eqnarray}
Usually in RSH, the $\omega$ is fixed as a constant (average value optimized for some reference data), a system-dependent constant (optimized for a given system), or even a position-dependent parameter. In the latter case, it was considered~\cite{Krukau2008Hybrid,Toulouse2004Short,Pollet2002Combining,Brutting2022Hybrid}
\begin{eqnarray}
\omega\sim \frac{1}{r_s}+\frac{s}{r_s}+\frac{s^2}{r_s} + \dots \;.    
\end{eqnarray}
%
with $r_{s}=(\frac{3}{4\pi n(\R)})^{1/3}$, $s=\frac{|\nabla n(\R)|}{2k_F n(\R)}$, $k_F=(3\pi^2 n(\R))^{1/3}$, and $n(\R)$ being the {\textcolor{black}{all electron total electron density}}. Hence, the Wigner-Seitz radius ($r_s$) alone should be, in principle, sufficient to define the screening parameter of a RSH functional, which is the main focus of this paper and is further elaborated in the text.
One may also argue that the density dependence of the range screening parameter was also proposed earlier~\cite{Baer2005Density}, but not established to use it in a practical way~\cite{Livshits2007well}.

Motivated by the above facts, we take a bit different approach to construct $\omega$. Particularly, we consider the long-wavelength limit of Eq.~\ref{eq:kernel2}, which is related to the exchange-correlation (XC) potential of the homogeneous electron gas (HEG) through the compressibility sum rule~\cite{Setsuo1987Statistical}
 \begin{equation}
  K_{xc}(q\to 0)=
  \frac{d^2}{dn^2}(n\epsilon_{xc}^{\text{LSDA}}(r_s,\zeta))~,
  \label{eq:kernel3}
 \end{equation}
 where \textcolor{black}{$\epsilon_{xc}^{LSDA}(r_s,\zeta)$} is the LSDA XC (PW91) energy per particle with $\zeta$ being the relative spin-polarization.
In $q\to 0$, from Eq.~\ref{eq:kernel2} and Eq.~\ref{eq:kernel3} results in,
\begin{equation}
\omega = \sqrt{-\frac{\pi}{K_{xc}(q\to 0)}}~.
\end{equation}

This form is further simplified in ref.~\cite{Jana2023Simple,Jana2025metagga,Jana2024accurate} by considering a fit to the exact form,
\begin{equation}
  \omega= \frac{a_1}{\langle r_{s} \rangle} + \frac{a_2 \langle r_{s} \rangle}{1 + a_3 \langle r_{s} \rangle^2}~,
 \label{mu-eff}
\end{equation}
where $a_1= 1.91718$, $a_2= -0.02817$, and $a_3=0.14954$. The local Seitz radius is given by $r_{s}=(\frac{3}{4\pi n})^{1/3}$ with $n=(n_{\uparrow}+n_{\downarrow})$ and 
$\langle r_s \rangle$ it's the average over volume (unit cell).
Although $\langle r_s \rangle$  is straightforward to calculate for bulk solids, tailored attention is required for finite systems such as atoms and molecules, where the Seitz radius diverges in the tail of the density. 
To this end, for this kind of system, we consider another definition of average $r_s$
\begin{equation}
\langle r_{s}\rangle=\frac{\int w({\bf{r}}) r_{s}({\bf{r}}) d^3r}{\int w({\bf{r}}) d^3r}\; .
\label{eq:rs-average}
\end{equation}
The $w(\mathbf{r})$ function is constructed in such a way as to catch the region where most of the electron density is localized (core and valence region). The same function is also used to define the volume for which we perform the averaging of $r_s$. Hence, we have defined this function as 
%
\begin{equation}
w(\mathbf{r}) = \mathrm{Erf}\left( \frac{n(\mathbf{r})}{n_{\text{c}}}\right),
\label{Eq:error}
\end{equation}
%
where Erf is the error function with the cutoff density threshold ($n_{\text{c}}$) defined as
\begin{equation}
n_{\text{c}}=\frac{n_{\text{th}}}{\int n(\mathbf{r}) \, d^3{\textit{r}}} \; .
\label{Eq:error2}
\end{equation}

{\textcolor{black}{
The cutoff density, $n_c$, and corresponding radius, $r_c = \left( \frac{3}{4\pi n_c} \right)^{1/3}$, are system- and size-dependent.  Our threshold selection of
$n_{th}=1.64\times10^{-2}$ e/bohr$^3$ 
provides consistent range-separation parameters ($\omega$) for charge-transfer molecules through ionization energy (IE) tuning (see Tables SI5-SI7). 
While exact matching $\omega_{eff}\simeq \omega_{IE}$ remains challenging due to their distinct physical origins, this $n_{th}$ value represents a balanced choice for accurate charge-transfer excitation energies. Figure S1 further validates our approach, demonstrating close agreement between $\omega_{eff}$ and $\omega_{IE}$ for linear acenes ($n = 2$-40) , poly(p-phenylenevinylene) molecules [(PPV)$_{n=1-8}$], and poly(p-phenyl)nitroaniline  [O$_2$N(Ph)$_{n=1-11}$NH$_2$] oligomers.  In contrast, alternative parameters  ($n_{th} = 10^{-1}~ e/bohr^{3}$ or $n_{th} = 10^{-3}~ e/bohr^{3}$) yield significantly different $\omega_{eff}$ values.}

{\textcolor{black}{Furthermore, within this scheme, the integral $\int n(\mathbf{r}') \, d^3{\textit{r}}'$ effectively captures size-dependent variations and delocalization differences in linear molecular chains (discussed subsequently). We also note that \Eq{Eq:error} tends to $1$ when
$\int n(\mathbf{r}') \, d^3{\textit{r}}'\to \infty$, recovering the bulk limit.
It is crucial to acknowledge that this methodology, like other tuned range-separated hybrid approaches, inherits the fundamental challenge of size inconsistency, a limitation extensively documented by Karolewski et al.~\cite{Karolewski2013Using}. This deficiency arises directly from the density integral formulation in Eq.~\ref{Eq:error2}, which introduces system-size dependence in parameter tuning. Thus the present approach, as well as GDD and the other
tuned RSHs will fail the calculation of molecular properties 
where size consistency plays a key role.}}


To illustrate the novelty of the construction, we present a comparative plot of the Wigner-Seitz radius for the CN molecule \textcolor{black}{( geometry from \cite{grabowski:2014:jcp})} and the product $r_s(\mathbf{r}) w(\mathbf{r})$ in the same panel in Fig.~\ref{fig1:rs}. As demonstrated, the quantity $r_s(\mathbf{r}) w(\mathbf{r})$ accurately reflects the behavior of $r_s(\mathbf{r})$ in regions of finite electron density, while it decays exponentially in the low-density tail regions. This exponential decay effectively captures the localization characteristics of the electron density, highlighting the ability of the constructed quantity to differentiate between the core and tail regions of the electronic distribution. 

\begin{figure*}
\centering   \includegraphics[width=\textwidth]{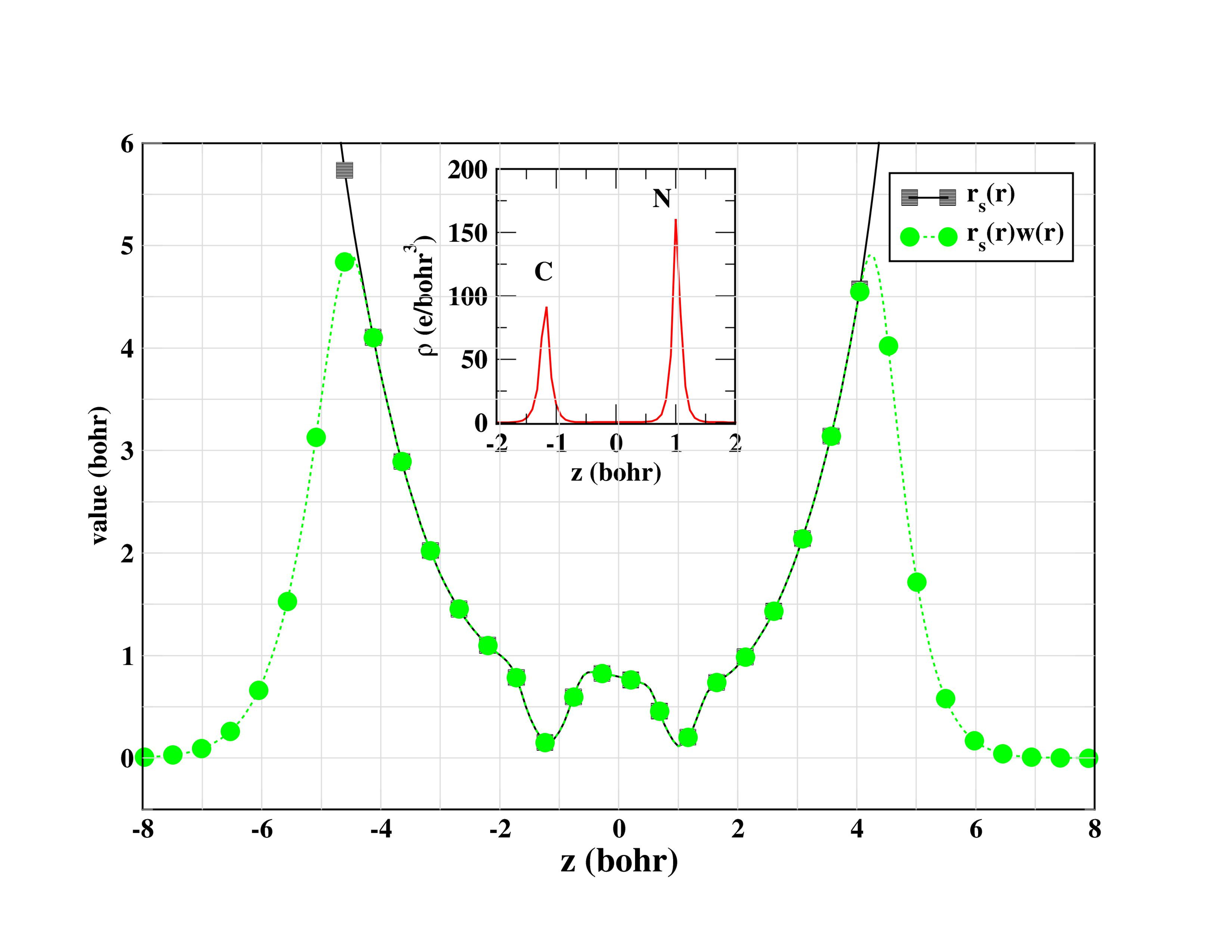}
\caption{The figure shows the electron density peaks for the CN molecule (inset) and 
the correct switching behavior of $r_s(\mathbf{r}) w(\mathbf{r})$, which effectively suppresses the exponential growth of $r_s(r)$.
}
    \label{fig1:rs}
\end{figure*}

To demonstrate in practice the efficiency of this methodology, we perform all calculations using the long-range corrected hybrid functional LC-$\omega$PBE, where the range-separation parameter $\omega$ is defined according to Eq. \ref{mu-eff}. 
The computational tools employed include PySCF~\cite{pyscf}, NWChem~\cite{nwchem}, and Q-Chem~\cite{qchem}. Specifically, PySCF is used to evaluate Eq.~\ref{eq:rs-average} and Eq.~\ref{mu-eff} according to the script deposited in the GitHub repository~\cite{repo}. {\textcolor{black}{The value of $\omega_{\text{eff}}$ does not need to be evaluated self-consistently. Instead, we recommend obtaining the electron density using the PBE exchange-correlation (XC) functional, and then using this density to construct $\omega_{\text{eff}}$. This part of the calculation is carried out using the cc-pVDZ basis set.  Importantly, we also observe that $\omega_{\text{eff}}$ varies only moderately with the choice of XC functional and basis set, and such variations have a negligible effect on the final results. We note that a similar scheme is also employed in the evaluation of $\omega_{\text{GDD}}$. 
}} Moreover, NWChem and Q-Chem are utilized to carry out all ground-state and TD-DFT calculations. The basis sets used in these calculations are specified either in the Supporting Information (SI) or within the captions of each table. 
The solid-state calculations are performed in the Vienna Ab initio Simulation Package (VASP)~\cite{vasp1,vasp2,vasp3,vasp4}. 
\begin{figure*}
\includegraphics[width=\textwidth]{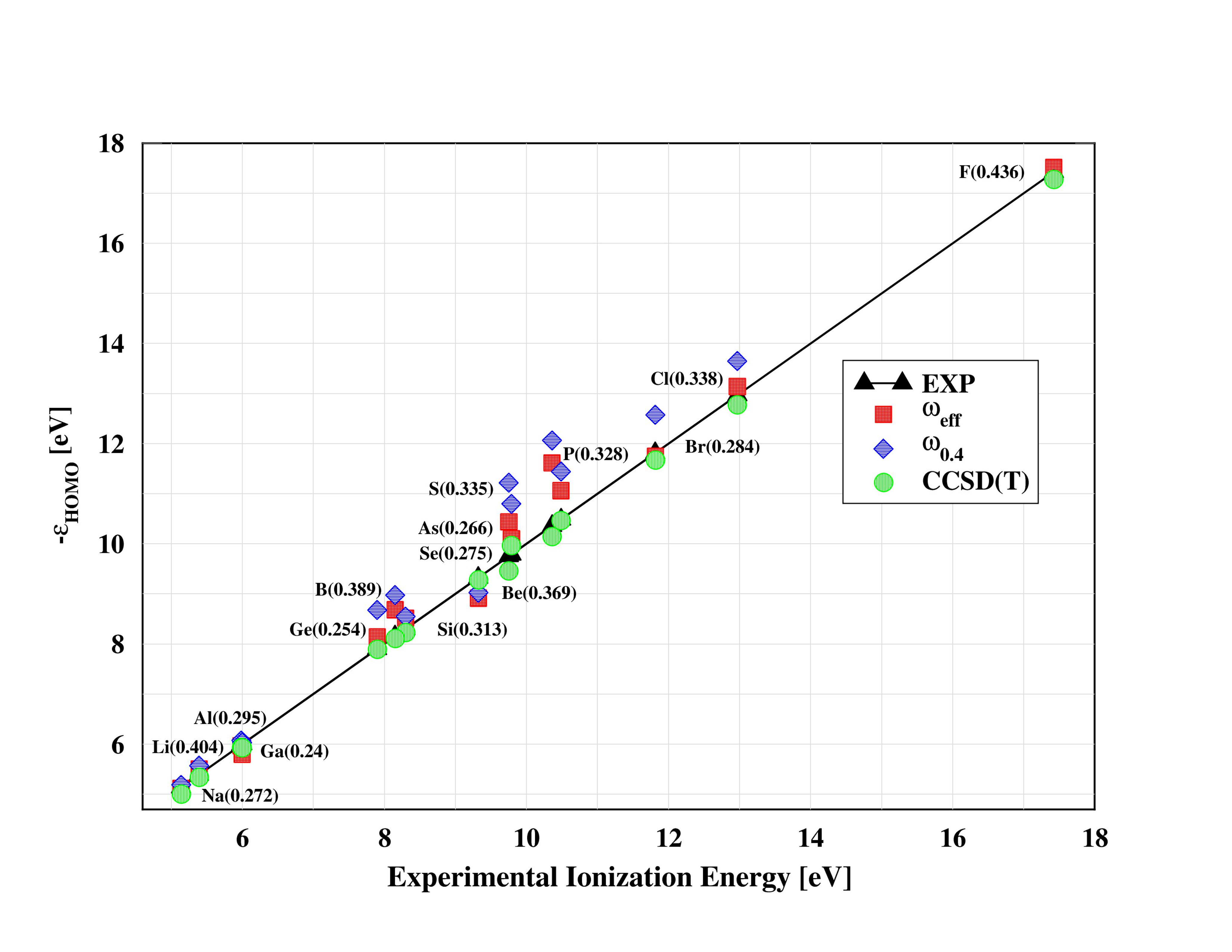}   
\caption{The ionization energies (IE) of various open-shell atoms calculated using the  LC-$\omega_X$PBE (x=0.4, eff) functional. The results are benchmarked against experimental data~\cite{Linstrom2001nist} and CCSD(T) reference values~\cite{Linstrom2001nist}. \textcolor{black}{The values within the parentheses correspond to $\omega_{eff}$.} The aug-cc-pVTZ basis set~\cite{aug-cc-pvtz} was utilized in all calculations. The data are compiled in Table SI1.}
    \label{fig1:atoms}
\end{figure*}


We first validate our approach by comparing the computed ionization energies of small atoms with available CCSD(T) and experimental values~\cite{Linstrom2001nist}, as shown in Fig.~\ref{fig1:atoms}. \textcolor{black}{We would like to refer to our newly tuned range separated parameter as $\omega_{eff}$.} The ionization energies computed from HOMO energies for LC-$\omega_{eff}$PBE value show good agreement with both reference datasets (MAE = 0.33 eV) outperforming the fix \textcolor{black}{($\omega = 0.4$),} LC-$\omega_{0.4}$PBE variant (MAE = 0.61 eV). Overall, the results are well balanced, with most values closely aligned along the diagonal line. One can also note how the $\omega_{eff}$ values (reported in first column of Table SI1) differentiate between the system varying between $\omega_{eff} \in (0.22 $-$ 0.42)$.

\begin{table*}[htbp]
\centering
\caption{TD-DFT charge-transfer excitation energies calculated using the  LC-$\omega_X$PBE (x=IE, eff) functional. The ionization energy (IE)-tuned range-separation parameter, $\omega$, and the excitation energies are taken from ref.~\cite{Mandal2025Simplified}. The theoretically best estimated (TBE) values (last column) taken as reference excitation energy are from  Ref.~\cite{loos2021}. All calculations employ the def2-TZVPD\cite{def2-TZVPD} basis set. The best results are highlighted in bold.}
\label{tab:ct-table}
\begin{tabular}{l cc cc c}
\toprule
Molecule & \multicolumn{2}{c}{$\omega$ (bohr$^{-1}$)} & \multicolumn{2}{c}{LC-$\omega$PBE (eV)} & E$_{\text{ref}}$ (eV) \\
\cmidrule(lr){2-3} \cmidrule(lr){4-5}
 & $\omega_{IE}$ & $\omega_{eff}$ & $\omega_{IE}$ & $\omega_{eff}$ & \\
\midrule
Aminobenzonitrile       & 0.293 & 0.275 & 5.38 & \textbf{5.36} & 5.26 \\
Aniline                 & 0.305 & 0.280 & \textbf{5.85} & 6.04         & 5.87 \\
Azulene                 & 0.241 & 0.266 & 3.72 & \textbf{3.91} & 3.89 \\
                        & 0.241 & 0.266 & \textbf{4.68} & \textbf{4.68}        & 4.55 \\
Benzonitrile            & 0.297 & 0.280 & \textbf{6.69} & 6.65         & 7.10 \\
Benzothiadiazole        & 0.443 & 0.269 & \textbf{4.43} & 4.47         & 4.37 \\
Dimethylaniline         & 0.266 & 0.266 & \textbf{4.59} &\textbf{4.59}         & 4.47 \\
                        & 0.266 & 0.266 & \textbf{5.44} & \textbf{5.44}         & 5.54 \\
Nitroaniline            & 0.284 & 0.274 & 4.44 & \textbf{4.59} & 4.57 \\
Nitrodimethylaniline    & 0.248 & 0.261 & 4.13 & \textbf{4.35} & 4.28 \\
Phthalazine             & 0.275 & 0.272 & 3.78 & \textbf{3.82} & 3.93 \\
                        & 0.275 & 0.272 & 4.35& \textbf{4.30}         & 4.34 \\
Quinoxaline             & 0.341 & 0.268 & 4.78 & \textbf{4.74} & 4.74 \\
                        & 0.341 & 0.268 & \textbf{6.02} & 6.13         & 5.75 \\
                        & 0.341 & 0.265 & \textbf{6.43} & 6.22         & 6.33 \\
Twisted DMABN           & 0.279 & 0.260 & \textbf{3.81} & 3.78         & 4.17 \\
                        & 0.279 & 0.260 & 5.17 & \textbf{4.91} & 4.84 \\
Dipeptide               & 0.325 & 0.264 & 7.99 & \textbf{8.22} & 8.15 \\
$\beta$-dipeptide       & 0.296 & 0.258 & 8.00 & \textbf{8.48} & 8.51 \\
                        & 0.296 & 0.258 & 9.28 & \textbf{8.79} & 8.90 \\
$N$-phenylpyrrole       & 0.464 & 0.261 & 5.72 & \textbf{5.53} & 5.53 \\
                        & 0.464 & 0.261 & 6.65 & \textbf{6.15} & 6.04 \\
DMABN                   & 0.257 & 0.263 & \textbf{4.89} & 5.02         & 4.94 \\

\bottomrule
MAE [eV] &&&0.22$^a$&{\textbf{0.12}} \\
\bottomrule
\end{tabular}
\begin{flushleft}
\footnotesize
$^a$ TBE used as ref.~\cite{loos2021} and MAE recalculated.\\
\end{flushleft}

\end{table*}

Next, we compare the \textcolor{black}{IE} tuning strategy 
with the present approach for describing charge-transfer (CT) excitations within the TD-DFT framework. The computed excitation energies are summarized in Table~\ref{tab:ct-table}, \textcolor{black}{(the geometries are extracted from Ref. \cite{Mandal2025Simplified})}, along with their corresponding range-separation parameters ($\omega$ values). Notably, the effective tuning parameter, $\omega_{\text{eff}}$, remains nearly constant across the set of studied molecules, a behavior also reported for the GDD approach in ref.~\cite{Mandal2025Simplified}.

The excitation energies obtained using the LC-$\omega_{\text{eff}}$PBE functional show strong agreement with those from the \textcolor{black}{IE-based} tuning method, indicating that the $\omega_{eff}$ values effectively captures long-range electron-hole interactions characteristic for CT states.

Remarkable improvements are observed for certain systems when using the LC-$\omega_{\text{eff}}$PBE functional, particularly for nitrodimethylaniline, azulene, nitroaniline, twisted DMABN, dipeptide, and $\beta$-dipeptide, where excitation energies align more closely with reference values. 
 \textcolor{black}{Overall, our method demonstrates superior accuracy with an MAE of 0.12 eV, substantially outperforming the standard IE-tuning strategy, which exhibits an MAE of 0.22 eV}. Additionally, employing $\omega_{\text{eff}}$ significantly enhances prediction accuracy over the GDD-derived $\omega_{GDD}$ values, which display a larger MAE of 0.19 eV\cite{Mandal2025Simplified}.
 \textcolor{black}{Given that $\omega_{0.3}$ approximates the mean of $\omega_{IE}$ and $\omega_{eff}$, we provide LC-$\omega_{0.3}$PBE data in Table SI2 for reference. This functional similarly produces a 0.19 eV MAE, matching $\omega_{GDD}$'s performance.} These findings highlight the $\omega_{\text{eff}}$ parameter's predictability and scalability for forecasting CT excitations in a variety of molecular systems.

\begin{table*}
\centering
\caption{TD-DFT singlet excitation energies within TDA approximations for several open-shell molecules calculated using the  LC-$\omega_X$PBE (x=IE, eff) functional.
The IE-tuned reference results are taken from Ref.~\cite{Mandal2025Simplified}. The theoretically best estimated values are taken from Ref.~\cite{open_data}. All calculations employ the def2-TZVPD basis set.}
\label{tab:transitions-open}
\begin{tabular}{@{}l c cc cc c@{}}
\toprule
Molecule & Transition & \multicolumn{2}{c}{$\boldsymbol{\omega}$ (bohr$^{-1}$)} & \multicolumn{2}{c}{LC-$\omega$PBE (eV)} & E$_\text{ref}$ (eV) \\
\cmidrule(lr){3-4} \cmidrule(lr){5-6}
 &  & $\omega_{IE}$ & $\omega_{eff}$ & $\omega_{IE}$ & $\omega_{eff}$ & \\
\midrule
BeF       & $^\circ\!\pi$                  & 0.496 & 0.290 & 4.20 & 4.18 & 4.13 \\
BH\textsubscript{2} & $^\circ$B$_1$          & 0.482 & 0.366 & 1.29 & 1.28 & 1.18 \\
CN        & $^\circ\!\pi$        & --    & 0.367 & --   & 1.47 & 1.33 \\
HCF       & $^1\!A^{\prime\prime}$          & 0.468 & 0.340 & 2.37 & 2.36 & 2.49 \\
NH\textsubscript{2} & $^2\!A_1$             & 0.659 & 0.353 & 2.02 & 2.11 & 2.11 \\
NO        & $^2\!\Sigma^+$                  & 0.600 & 0.382 & 5.74 & 6.08 & 6.12 \\
OH        & $^2\!\Sigma^+$                  & 1.547 & 0.371 & 4.77 & 4.02& 4.09 \\
NCO       & $^2\!\Sigma^+$                  & 1.515 & 0.349 & 3.75 & 3.16 & 2.89 \\

\bottomrule
MAE[eV]&&&&0.29$^a$&{\bf{0.10}}&\\
\bottomrule
\end{tabular}
\begin{flushleft}
\footnotesize
$^a$ We consider as reference the TBE values from ref.~\cite{open_data} and recalculate MAE. Detailed Error statistics available in Table SI3.\\
\end{flushleft}
\end{table*}

\begin{figure*}
    \centering   
    \includegraphics[width=17.4 cm, height = 15 cm]{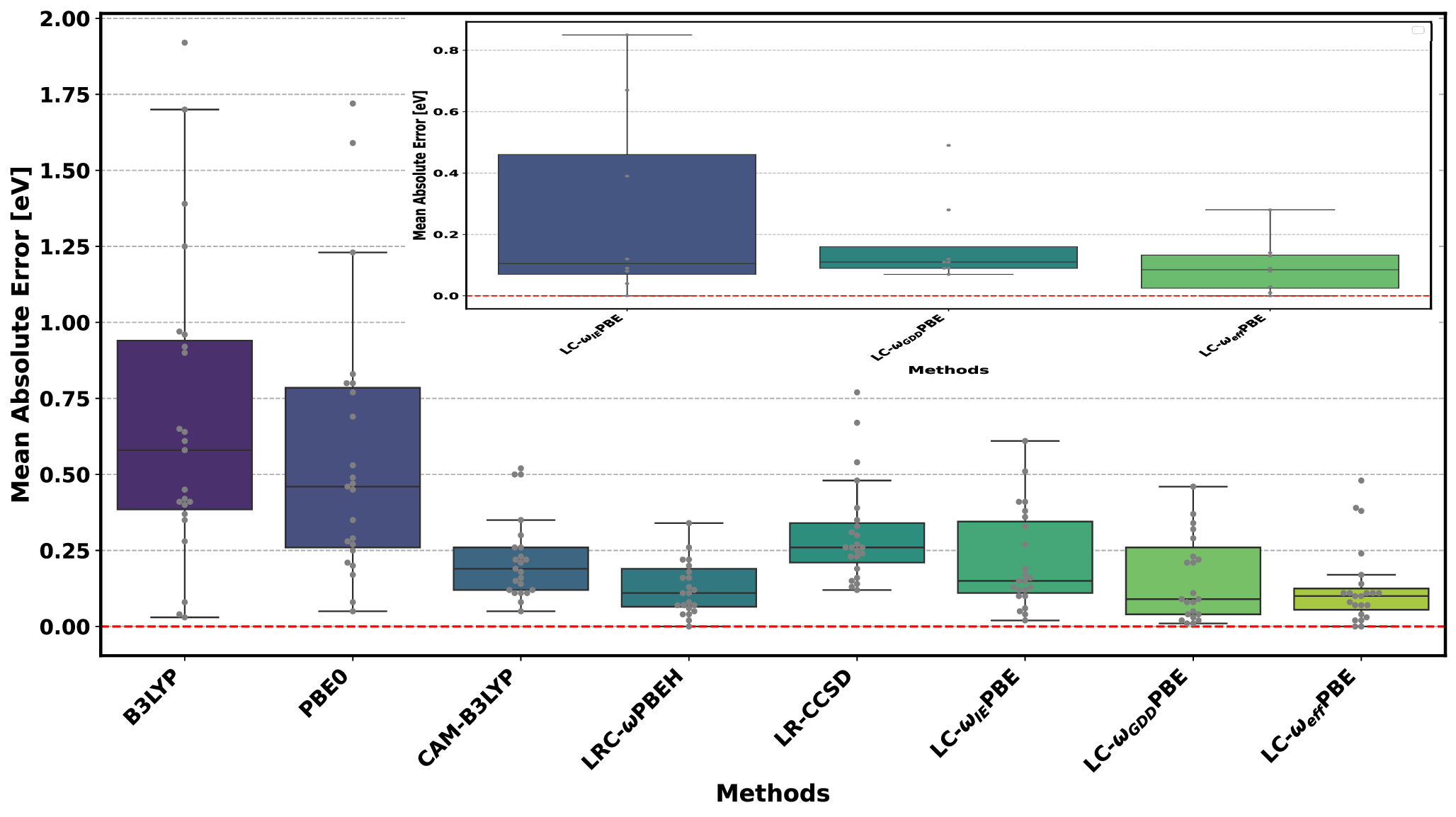}
   
    \caption{Box plots for the mean absolute error (MAE) of the benchmark test set of Table~\ref{tab:ct-table}. The LC-$\omega_{IE}$PBE, B3LYP, PBE0, CAM-B3LYP, LRC-$\omega$PBEH are taken from Ref.~\cite{Mandal2025Simplified} whereas the  LR-CCSD  values are taken from Ref.\cite{loos2021}. For Table~\ref{tab:transitions-open}, the box plot is given in an inset for three methods. The data is available in Tables SI2 and SI3, respectively.}
    \label{fig1:error1}
\end{figure*}

\begin{figure*}
     \includegraphics[width=17 cm, height = 17 cm]{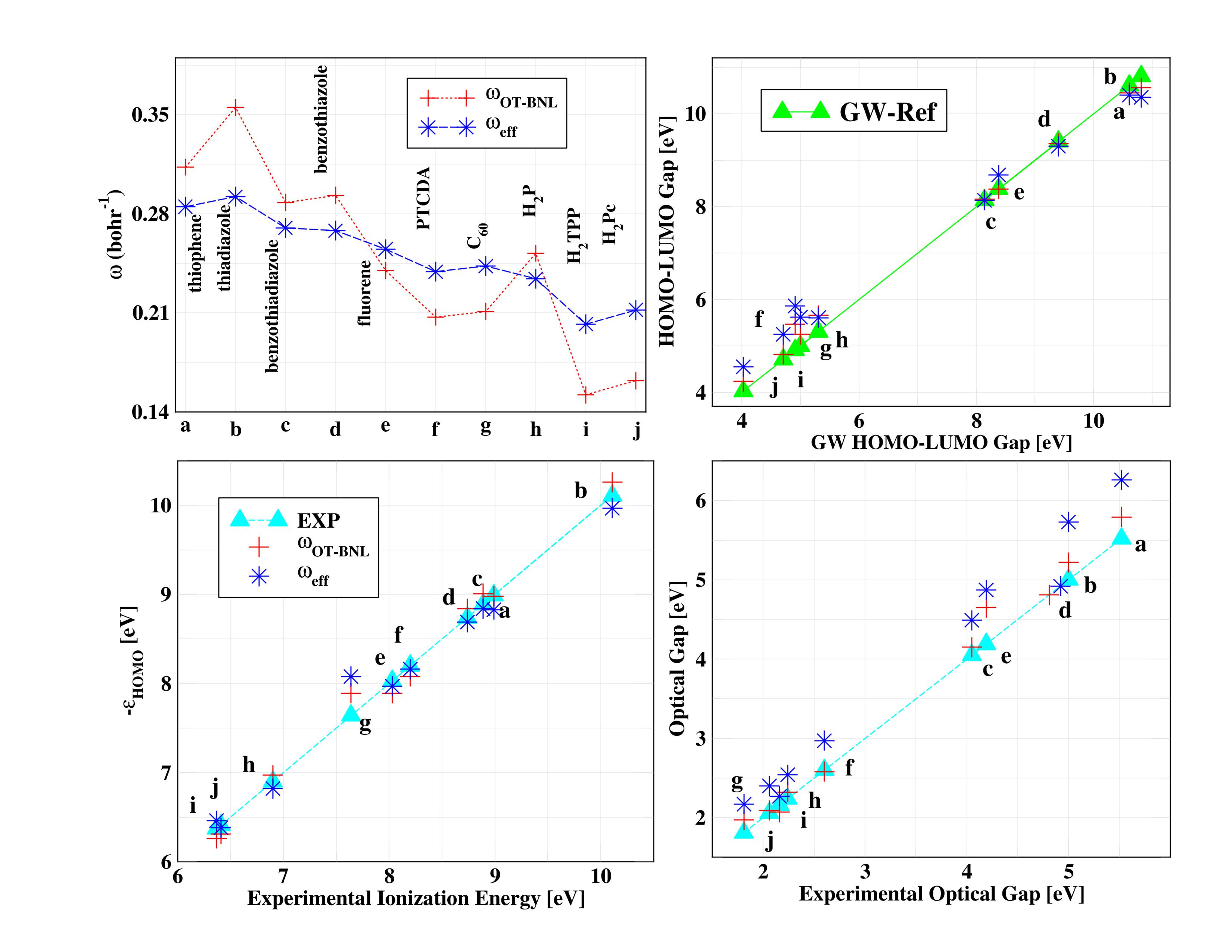}    \caption{Variation of $\omega$, HOMO energies, HOMO-LUMO gap, and optical gaps of relevant organic \textcolor{black}{photo-voltaic} (OPV) molecules. The LC$-\omega_{OT-BNL}$PBE, $GW$, and experimental results are taken from ref.~\cite{Refaely2011Fundamental}. Calculations are performed using the cc-pVDZ basis set. Comprehensive data is available in Table SI4. Note that for LC$-\omega_{OT-BNL}$PBE the screening parameter is optimized with respect to both $N$ {\textcolor{black}{(HOMO)}} and $N+1$ {\textcolor{black}{(LUMO)}} molecular orbitals.}
    \label{fig1:photovoltic}
\end{figure*}


\begin{table*}
\centering
\caption{\label{tab:acene-chain} {\textcolor{black}{Table presents the absolute deviations from experimental values for vertical excitation energies in linear acene rings, benchmarked against experimental data from Ref.~\cite{la}. We also present mean absolute error (MAE)}}.  
Results for the LC-$\omega_X$PBE (x=GDD, IE) are included from Ref.~\cite{Mandal2025Simplified}. All computations employed the def2-TZVPD basis set.}
\resizebox{0.8\textwidth}{!}{

\begin{tabular}{l c c c c c c c c c}
\toprule
Molecule & Transition & Exp & \multicolumn{3}{c}{TD-LC-$\omega_X$PBE Error (eV)} \\
\cmidrule(lr){4-6}
 & & &  $\omega_{GDD}$ & $\omega_{IE}$ &$\omega_{eff}$ \\
\midrule
Naphthalene & \(^{1}\textrm{L}_{a}\) & 4.66 & 0.09 & 0.01 &0.02 \\
Naphthalene & \(^{1}\textrm{L}_{b}\) & 4.13 & 0.45 & 0.40 &0.40  \\
Anthracene & \(^{1}\textrm{L}_{a}\) & 3.60 & 0.03 & 0.11 &0.10 \\
Anthracene & \(^{1}\textrm{L}_{b}\) & 3.64 & 0.41 & 0.30 &0.36\\
Tetracene & \(^{1}\textrm{L}_{a}\) & 2.88 & 0.07 & 0.18 &0.10 \\
Tetracene & \(^{1}\textrm{L}_{b}\) & 3.39 & 0.34 & 0.18 &0.30 \\
Pentacene & \(^{1}\textrm{L}_{a}\) & 2.37 &  0.05 & 0.16 &0.08 \\
Pentacene & \(^{1}\textrm{L}_{b}\) & 3.12 &  0.41 & 0.21 &0.37\\
Hexacene & \(^{1}\textrm{L}_{a}\) & 2.02 & 0.03 & 0.14 &0.06 \\
Hexacene & \(^{1}\textrm{L}_{b}\) & 2.87 &  0.16 & 0.02 &0.10\\
\midrule
\multicolumn{3}{l}{MAE(eV)}  & 0.204 & 0.171 &0.189\\
\bottomrule
\end{tabular}}
\end{table*}

\begin{figure*}
    \centering    \includegraphics[width=1.1\textwidth]{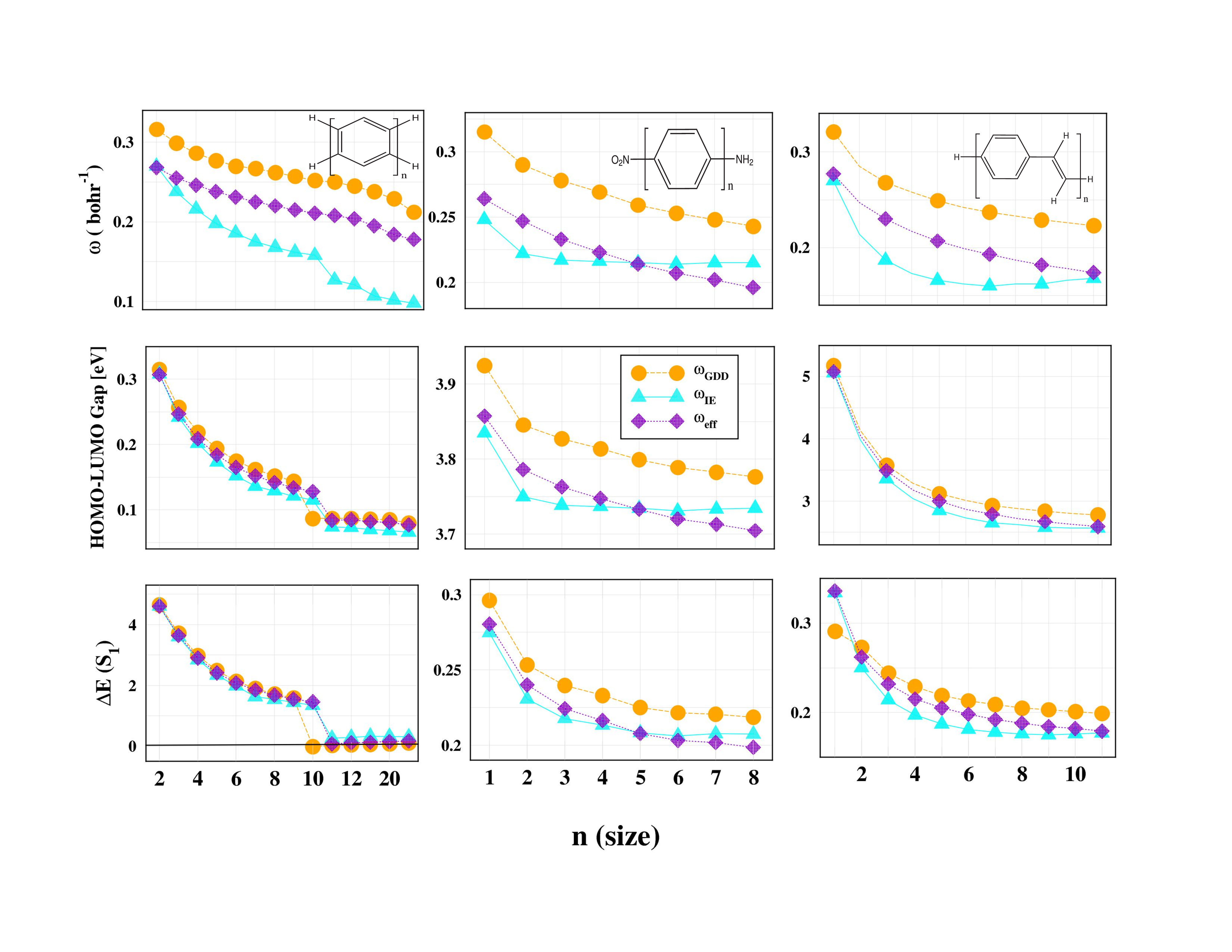}    
    \caption{Range-separation parameter ($\omega$), HOMO-LUMO gap, and singlet excitation energies for (from left to right): linear acenes ($n = 2$-40), poly(p-phenylenevinylene) molecules [(PPV)$_{n=1-8}$] and poly(p-phenyl)nitroaniline oligomers [O$_2$N(Ph)$_{n=1-11}$NH$_2$]. The $\omega_{GDD}$ and $\omega_{IE}$ values (along with their corresponding long-range corrected functional results) are taken from Ref.~\cite{Mandal2025Simplified}. All calculations were performed using the def2-ma-SVP basis set\cite{def2-ma-svp}. Comprehensive data is available in the \textcolor{black}{Tables} SI5, SI6, and SI7. }
    \label{fig1:large-molecules}
\end{figure*}

Figure~\ref{fig1:error1} presents a comparative analysis of various methods for charge-transfer (CT) excitations across the benchmark test set listed in Table~\ref{tab:ct-table}.
As shown, the LC-$\omega_{\text{eff}}$PBE functional yields the lowest MAE within different LC-$\omega$PBE variants,
demonstrating superior accuracy relative to several widely used traditional methods. In particular, LC-$\omega_{\text{eff}}$PBE outperforms B3LYP (MAE = 0.68 eV), PBE0 (MAE = 0.56 eV), CAM-B3LYP (MAE = 0.22 eV), and LRC-$\omega$PBEH (MAE = 0.17 eV) in this context. We have also extensively compared our results with linear-response coupled-cluster singles and doubles (LR-CCSD) results, which yield here an MAE of about 0.31 eV\cite{loos2021}. It is readily apparent that the poorest options for capturing CT excitations would be PBE0 and B3LYP functionals, which can be related to their wrong asymptotic decay of exchange-correlation potential\cite{SmigaPot}. Notably, despite LRC-$\omega$PBEH\cite{lrc_w_pbeh,lrc_w_pbeh_1,lrc_w_pbeh_2} being explicitly tailored for systems with charge-separation characteristics, the LC-$\omega_{eff}$PBE functional demonstrates superior performance even surpassing it in accuracy. Thus, indicating its robustness and reliability for modeling CT excitations. A comprehensive overview of the data is available in Table SI2.

Next, the performance of the present method is benchmarked for singlet excitations of small open-shell molecules \textcolor{black}{(the geometries are acquired from \cite{Mandal2025Simplified})}, which are known to be challenging for the IE-tuning approach. As shown in Table~\ref{tab:transitions-open}, optimized $\omega_{IE}$ values are not always obtainable. For example, no IE-tuned value could be found for the CN radical, where $\omega_{IE} \to \infty$~\cite{Mandal2025Simplified}. In contrast, the $\omega_{\text{eff}}$ values
are more systematic and exhibit significantly less variation across different systems, compared to IE-tuning. For radicals such as OH and NCO, singlet excitation energies are generally overestimated by LC-$\omega_{{\text{IE}}}$PBE relative to LC-$\omega_{\text{eff}}$PBE. Overall, LC-$\omega_{\text{eff}}$PBE yields a MAE of 0.10 eV, which is substantially lower than that obtained from both $\omega_{{\text{IE}}}$ and $\omega_{{\text{GDD}}}$ methods~\cite{Mandal2025Simplified}. Notably, the IE and GDD approaches exhibit significant change
of $\omega$, consequently inflating the calculated singlet excitation energies. Crucially, LC-$\omega_{eff}$PBE demonstrates robust performance not only for closed-shell systems but also achieves accurate predictions for open-shell charge-transfer (CT) excitations, establishing its versatility across diverse systems.

Next, we turn our attention to organic photovoltaic (OPV) materials. 
Optimized range-separated hybrid functionals have proven highly effective in reliably predicting the electronic properties of these materials, offering precision in modeling critical charge-transfer and excitation behaviors~\cite{Refaely2011Fundamental}. These organic systems are typically large in size, which makes conventional optimal-tuning (\textcolor{black}{tuned using both $N$ \textcolor{black}{(HOMO)} and $N+1$ \textcolor{black}{(LUMO)} molecular orbitals energies})
procedures computationally expensive. In this context, the use of a simplified yet effective, range-separation parameter such as $\omega_{\text{eff}}$ may offers a highly practical alternative without sacrificing accuracy.

For benchmarking purposes, we consider the same representative set of OPV molecules as studied in Ref.~\cite{Refaely2011Fundamental}.  \textcolor{black}{These geometries were obtained from.\cite{Refaely2011Fundamental}}. The performance of the LC-$\omega_{\text{eff}}$PBE functional is illustrated in Fig. \ref{fig1:photovoltic}. An initial analysis of the $\omega_{\text{eff}}$ values reveals good agreement with optimal tuned values for most systems. Notable deviations can be seen, e.g., thiadiazole, PTCDA or C$_{60}$. However, these deviations have minimal impact on the calculated electronic and excitonic properties. The HOMO energies computed with LC-$\omega_{\text{eff}}$PBE align closely with the reference diagonal, indicating accurate predictions. A similar trend is observed for the HOMO-LUMO gaps, with LC-$\omega_{\text{eff}}$PBE values showing excellent agreement with benchmark $GW$ results. Furthermore, the optical gaps calculated using LC-$\omega_{\text{eff}}$PBE match very closely with the optimally tuned \textcolor{black}{Baer, Neuhauser, and Livshits} (OT-BNL) values reported in Ref.~\cite{Refaely2011Fundamental}, reaffirming the reliability of the present approach.

Overall, these results highlight the effectiveness and practical applicability of LC-$\omega_{\text{eff}}$PBE for large OPV molecules, offering a computationally efficient yet accurate alternative to traditional IP- and/or gap-tuning methods.

In Table~\ref{tab:acene-chain} 
, we present a mean error analysis of the vertical excitation energies of linear acene systems \textcolor{black}{(the geometries are taken from Ref. \cite{Mandal2025Simplified})}, benchmarked against experimental reference values \cite{la}. The excitation comprises two distinct transition states: \textbf{L\textsubscript{a}}, corresponding to the HOMO $\rightarrow$ LUMO transition, and \textbf{L\textsubscript{b}}, which involves either a HOMO $\rightarrow$ LUMO+1 or HOMO$-$1 $\rightarrow$ LUMO transition. These states exhibit unique properties: the \textbf{L\textsubscript{a}} state displays significant ionic character in its wavefunction, while the \textbf{L\textsubscript{b}} state is predominantly covalent, resembling the ground state's characteristics\cite{la}. Notably, the simplified approach employing $\omega_{eff}$ achieves a MAE of 0.189 eV, which is comparable to the MAE of 0.171 eV obtained using $\omega_{IE}$. These results demonstrate that the streamlined methodology preserves accuracy in describing electronically complex linear conjugated systems. Furthermore, as shown in Table SI7, the value of $\omega_{eff}$ lies between those of $\omega_{GDD}$ and $\omega_{IE}$, supporting the observed performance trends.

One of the most important features of tuned RSH functionals is the size dependence of the range-separation parameter ($\omega$) with respect to the number of repeat units in conjugated systems such as polyenes and alkane chains. This characteristic has been extensively studied in previous works~\cite{Korzdorfer2011long, Korzdorfer2014Organic}. In such systems, time-dependent density functional theory (TD-DFT) calculations employing either global hybrid functionals or fixed-$\omega$ RSHs often suffer from delocalization or localization errors, leading to inaccurate electronic and optical properties. These challenges become even more pronounced in extended systems and nanoscale materials, as discussed in Ref.~\cite{Wang2016Hybrid}.

To validate the performance and size-scaling behavior of our present approach, we examine three prototypical classes of linear conjugated molecules: (i) linear acenes with $n=2$-40 benzene rings, (ii) poly(p-phenylenevinylene) oligomers [(PPV)$_{n=1-8}$], and (iii) poly(p-phenyl)nitroaniline chains [O$_2$N(Ph)$_{n=1-11}$NH$_2$]. These structures were previously analyzed in the context of the GDD approximation in ref.~\cite{Mandal2025Simplified}, \textcolor{black}{(the geometries are extracted from Ref. \cite{Mandal2025Simplified})}.

As shown in Fig.~\ref{fig1:large-molecules}, the $\omega_{\text{eff}}$ values predicted by our method closely follow the trends observed for the values tuned with the ionization energy ($\omega_{IE}$), and are consistently lower than those obtained from the GDD approach ($\omega_{GDD}$). This is true across all three classes of molecules, suggesting that $\omega_{\text{eff}}$, \textcolor{black}{has a distinct cutoff, which is introduced in \Eq{Eq:error2} that} captures the correct size dependence associated with electronic delocalization \textcolor{black}{which it is related to.}

In terms of electronic properties, both the HOMO-LUMO gaps and singlet excitation energies exhibit a monotonic decrease with increasing chain length, which is in line with physical expectations. However, the results from LC-$\omega_{IE}$PBE show a more pronounced decay compared to LC-$\omega_{GDD}$PBE. The LC-$\omega_{\text{eff}}$PBE results, on the other hand, follow the trend of LC-$\omega_{IE}$PBE more closely, indicating that the present method better mitigates delocalization errors in long-chain systems. A particularly notable case arises for linear acenes with $n=10$, where the LC-$\omega_{GDD}$PBE predicts a negative excitation energy for one singlet state, an unphysical artifact that does not appear in either LC-$\omega_{\text{eff}}$PBE or LC-$\omega_{IE}$PBE, highlighting the improved stability of the $\omega_{\text{eff}}$-based approach.
\textcolor{black}{A discontinuity is evident when comparing acenes with $n=10$ and $n=11$ rings, attributable to a sudden change in the Kohn-Sham gap. This observation implies the emergence of an open-shell biradicaloid singlet ground state in longer acenes.\cite{Mandal2025Simplified}}
For the \textcolor{black}{$(PPV)_{n=1-8}$} oligomers, LC-$\omega_{\text{GDD}}$PBE significantly overestimates the HOMO-LUMO gap compared to both LC-$\omega_{IE}$PBE and LC-$\omega_{\text{eff}}$PBE, with the latter two in closer mutual agreement and better alignment with expected physical behavior. Similarly, for the poly(p-phenyl)nitroaniline chains, $\omega_{\text{eff}}$ again follows the size trend of $\omega_{IE}$ more accurately than $\omega_{GDD}$. Although the differences in HOMO-LUMO gaps and excitation energies are less dramatic for this system, the consistency of the $\omega_{\text{eff}}$ trend supports the robustness of the present method. {\textcolor{black}{We also want to mention that for longer molecular chains, Wannier-optimized tuning may also be necessary for longer molecular chains}}


\begingroup
\renewcommand{\arraystretch}{1.2}
\begin{table}[h!]
\centering
\caption{Screening parameters for bulk and monolayer solids. All values are in Bohr$^ {-1}$.}
\label{tab:solids}

\begin{tabular}{lccccc}
\toprule
\textbf{Material} & $\mu^{a}$ & $\mu_{eff}^{fit~b}$ & $\mu_{WS}^{b}$ & $\mu_{TF}^{b}$ & $\omega_{eff}$ \\
\midrule
\multicolumn{6}{c}{\textit{Periodic bulk solids}} \\
Ar        & 0.74 & 0.54  & 0.52 & 0.56 & 0.51 \\
C         & 0.90 & 1.24  & 0.76 & 0.68 & 1.25
\\
Ge        & 0.62 & 0.79  & 0.45 & 0.52 & 0.72
\\
Si        & 0.65 & 0.85  & 0.50 & 0.55 & 0.82
\\
\midrule
\multicolumn{6}{c}{\textit{Periodic monolayers}} \\
Graphene ($c = 8$\,\AA)   & --   & 0.171 & --   & --   & 0.448
\\
$h$BN ($c = 20$\,\AA)     & --   & 0.072 & --   & --   & 0.369\\
$h$BN ($c = 22$\,\AA)& --   & -- & --   & --   &0.363
\\
\midrule
\multicolumn{6}{c}{\textit{Surfaces}$^c$} \\
Si(111)-(2$\times$1)&--&--&--&--&0.684\\
Ge(111)-(2$\times$1)&--&--&--&--&0.674\\
\bottomrule
\end{tabular}
\begin{flushleft}
\footnotesize
$^{a}$ Values taken from Ref.~\cite{WeiGiaRigPas2018}. \\
$^{b}$ Values reported in Ref.~\cite{Jana2023Simple}. \\
$^{c}$ Structures are generated from Ref.~\cite{Pandey1981New}.
\end{flushleft}
\end{table}
\endgroup

\begingroup
\renewcommand{\arraystretch}{1.2}
\begin{table*}
\centering
\caption{Screening parameters, band gaps, and position of the optical transition from TD-DFT calculations for molecular crystals. \textcolor{black}{Bracketed} values are $\omega_{eff}$ (in Bohr$^{-1}$) obtained from their respective gas-phases. {\textcolor{black}{Here, all calculations of the SRSH functional are performed with the dielectric constant ($\epsilon$) supplied in the supporting information of Ref.~\cite{Manna2018Quantitative}}}.}
\label{tab:mol-crystal}
\resizebox{1\textwidth}{!}{

\begin{tabular}{cccccccccccccc}
\toprule
\textbf{Material} & $\mu^a$ & $\mu_{TF}^b$ & $\omega^{OT-SRSH}$$^d$ & $\omega_{eff}$ 
& $E_g^g$ & $E_g^h$ & $G_0W_0$@PBE$^g$ 
& TD-SRSH$^g$ & TD-SRSH($\omega_{eff}$) & BSE$^g$ \\
\midrule
&&&& & \multicolumn{3}{c}{\textit{band gap}} & \multicolumn{3}{c}{\textit{optical gap}} \\
\cline{6-8} \cline{9-11}
NH$_3$  & 0.53$^e$ & 0.57$^e$ & 0.375 & 0.599 (0.364) & 7.9  & 8.3  & 7.7 & 7.1 & 7.8 & 7.1 \\
CO$_2$  & -- & -- & 0.405 & 0.546 (0.348)& 11.2 & 11.8 & 11.2 & 10.7 & 11.1 & 10.8 \\
\bottomrule
\end{tabular}}
\begin{flushleft}
\footnotesize
$^a$ From Ref.~\cite{WeiGiaRigPas2018}. \\
$^b$ From Ref.~\cite{Jana2023Simple}. \\
$^d$ Obtained from the gas phase of molecular crystals~\cite{Manna2018Quantitative}. \\
$^e$ From Ref.~\cite{Skone2016Nonempirical}. \\
$^g$ From Ref.~\cite{Manna2018Quantitative}. \\
$^h$ SRSH with $\omega_{\text{eff}}$.
\end{flushleft}
\end{table*}
\endgroup

%
A natural question arises: Is the present formalism equally applicable to solid-state systems? To address this, we note that the proposed scheme can be effectively extended to bulk solids. In this context, the bulk-limit behavior of the screening parameter must be consistent with that described in Ref.~\cite{Jana2023Simple}. To achieve this, we recommend modifying Eq.~\ref{Eq:error2} in a manner similar to Refs.~\cite{Rauch2020Accurate,Tran2021Bandgap}, using \( n_c = n_{\text{th}} = 6.96 \times 10^{-4}~e/\text{bohr}^3 \), which corresponds to a Wigner-Seitz radius of \( r_c = r_{\text{th}} = 7~\text{bohr} \), a suitable cutoff value for most bulk solids. However, in the case of solids, one must also account for dielectric-dependent effects~\cite{ZhengGovoniGalli2019} in conjunction with range-separated screening.

%
%

To illustrate the applicability of this form of screening parameter, in Table~\ref{tab:solids} and Table~\ref{tab:mol-crystal}, we show the $\omega$ values for a few bulk solids and molecular crystals. Our test consists of (i) periodic bulk ({\textcolor{black}{geometries are obtained from Ref.~\cite{Jana2023Simple}}}), {\textcolor{black}{(ii) 2D monolayers with various supercell heights or length perpendicular
to the 2D layers ($c$)}}, (iii) surfaces ({\textcolor{black}{geometries are obtained from Ref.~\cite{Pandey1981New}}}), and (iv) molecular crystals \textcolor{black}{(the geometries are obtained from Ref.\cite{Manna2018Quantitative})}. These are different kinds of solids that represent different physics of materials.

{\textcolor{black}{For comparison between different screening parameters, we consider the following: (i) the parameter $\mu$ fitted from the long-wavelength limit of the dielectric function~\cite{WeiGiaRigPas2018}, (ii) the effective screening parameter $\mu_{\text{eff}}^{\text{fit}}$ from Ref.~\cite{Jana2023Simple}, and (iii) $\mu$ estimated from the valence electron density ($n_v$), defined as the number of valence electrons per unit volume. Two forms of $\mu$ are proposed: 
\[
\mu_{\text{WS}} = \left( \frac{4\pi n_v}{3} \right)^{1/3} \quad \text{and} \quad \mu_{\text{TF}} = \left( \frac{3n_v}{\pi} \right)^{1/6},
\]
corresponding to the Wigner-Seitz (WS) and Thomas-Fermi (TF) screening models, respectively. Notably, the expressions for $\mu_{\text{eff}}^{\text{fit}}$, $\mu_{\text{WS}}$, and $\mu_{\text{TF}}$ involve the unit cell volume ($V_{cell}$) through $n_v$ or in the expression itself. In contrast, the effective frequency $\omega_{\text{eff}}$ proposed in this work differs in that contributions from regions of vanishing electron density are excluded. We argue that this approach is more general and applicable across a broader range of solid-state systems.}}

For bulk solids the results are closely matching with $\mu^{fit}_{eff}$ proposed in Ref.~\cite{Jana2023Simple}, clearly indicating the practical applicability of the present form. For 2D monolayers, $\omega_{eff}$ is not changing much with respect to different vacuum sizes ($c$) \textcolor{black}{(within periodic computational frameworks, 'vacuum size' defines the engineered empty-space dimension isolating structures (surfaces, slabs, etc.), crucially governing boundary-condition implementations.)}, which is very important as the present construction avoids the divergence of $r_s$. For 2D monolayers, $\mu_{TF}$, $\mu_{WS}$, and $\mu_{eff}^{fit}$ are not applicable {\textcolor{black}{as all these expression involves volume of the unit cell and the volume depends on $c$ values. Thus larger $c$ can give larger volume, which make $\mu_{TF}$, $\mu_{WS}$, and $\mu_{eff}^{fit}$ not useful in this case.}} As for example the application of $\mu_{eff}^{fit}$ for Graphene and $h$BN monolayers results very unphysical values, which are also tending to zero. Thus $\omega_{eff}$ is more general and robust (also remains almost fixed value with different vacuum size). For the Si(111) and Ge(111) surfaces, we also obtain reasonable $\omega_{eff}$. 

{\textcolor{black}{For molecular crystals we also tested $\omega_{eff}$ against $\omega$, tuned from optimally-tuned screened range-seperated hybrid (OT-SRSH). As noted from Table~\ref{tab:mol-crystal}, these values are slightly larger than OT-SRSH~\cite{Manna2018Quantitative}.}} This is not surprising as Ref.~\cite{Manna2018Quantitative} tuned the $\omega$ values from their respective gas phase, but not from bulk crystals. In Table~\ref{tab:mol-crystal} we also showed that $\omega_{eff}$ for gas phase matches quite close to that of $\omega_{OT-SRSH}$. Also, $\mu_{eff}^{fit}$ matches closely to $\mu$ and $\mu_{TF}$, indicating versatility of this method. For fundamental or KS gaps, we obtain a gap of 8.3 eV for NH$_3$ and 11.8 eV for CO$_2$ molecular crystals, which are quite close to $G_0W_0$ values. For molecular crystals, no tuning is applied for their respective finite systems. Thus, these adjustments ensure that the method remains both robust and physically meaningful when applied to solid-state environments. Also, note that $\omega_{eff}$ 
values are quite close to $\mu_{TF}$ or obtained from fitting with RPA data~\cite{Skone2016Nonempirical}. Finally, the obtained band gaps and positions of the first bright excitons, which are quite close to the benchmark \textcolor{black}{TD-SRSH, $G_0W_0@PBE$, and BSE} for various excited-state properties.      


Although this form is potentially insightful, particularly for extended systems with vacuum regions, such as two-dimensional materials, molecular crystals, or surfaces, this formulation (i.e., the dielectric-dependent functional development) requires further investigation for practical use. 
It is worth noting that for bulk solids, the $\Delta$SCF or IP-tuning or optimal-tuning procedures are generally not applicable because of the delocalization of orbitals~\cite{Dahvyd2021Band}; in this regard, the present method may offer advantages over the earlier proposed methods~\cite{SteiEiseHele2010,Refaely2012Quasiparticle,Skone2016Nonempirical}.

In conclusion, we have developed a simple and efficient single-shot approach to determine the range-separation parameter in long-range corrected hybrid functionals. The construction is based on a well-grounded, clear, and physically transparent framework.
Unlike conventional tuned range-separated hybrid methods, this novel approach achieves remarkable accuracy with significantly reduced computational cost across a wide variety of {\textcolor{black}{systems. This is important for modeling excitations in molecules}}. A key advantage of the present method is that the tuning behavior is derived solely from the electron density, making it
easily transferable and {\textcolor{black}{broadly} applicable. This represents a significant advancement toward achieving highly accurate results without requiring multiple tuning procedures. The application of the $\omega_{eff}$ tuning demonstrates superior performance in case of 
charge-transfer excitations, HOMO-LUMO gaps, and exciton energies.
Overall, this development broadens the applicability of range-separated hybrid functionals and opens new possibilities for interdisciplinary research, including the starting point of high-level methods~\cite{McKeon2022optimally}.

In future work, we will explore the application of this methodology in ground-state DFT and linear response TD-DFT calculations as well as its impact on functional, orbital, and density errors\cite{Aditi_orbital}.

{\textcolor{black}{\section*{Setup for VASP Calculations}
VASP calculations to evaluate $\omega_{\text{eff}}$ for periodic bulk solids are performed using the final electron density obtained from self-consistent calculations with the PBE exchange-correlation functional. A $\Gamma$-centered $15 \times 15 \times 15$ $\mathbf{k}$-point mesh and an energy cutoff of 600~eV are used.
For periodic monolayers and surface systems, a $\Gamma$-centered $15 \times 15 \times 1$ $\mathbf{k}$-point grid is employed, with the same energy cutoff of 600~eV. In the case of molecular crystals, a $\Gamma$-centered $8 \times 8 \times 8$ $\mathbf{k}$-point mesh and a higher energy cutoff of 800~eV are used to evaluate $\omega_{\text{eff}}$ (based on the PBE density) and to perform TD-SRSH calculations.}}



\section*{Data Availability}
All data supporting the findings of this study are available in the main text and the Supplementary Information section. The PySCF-based code for calculations of $\omega_{eff}$ values used in this work is publicly available in the repository~\cite{repo}.

\section*{Supporting Information}

The tables with raw data and additional figures supporting the analysis. 

\section*{Acknowledgements}
S.\'S. acknowledges the financial support from the National Science Centre, Poland (grant no. 2021/42/E/ST4/00096).
L.A.C. and F.D.S. acknowledge the financial support from ICSC - Centro Nazionale di Ricerca in High Performance Computing, Big Data and Quantum Computing, funded by the European Union - NextGenerationEU - PNRR.


\twocolumngrid

\bibliography{reference}
\bibliographystyle{plainurl}

\end{document}